\documentclass[12pt]{spieman}  
\usepackage{amsmath,amsfonts,amssymb}
\usepackage{graphicx}

\usepackage{tocloft}
\usepackage{lineno}
\usepackage[colorlinks=true, allcolors=blue]{hyperref}
\usepackage{booktabs}
\usepackage[section]{placeins}

\title{mach: ultrafast ultrasound beamforming}

\author[a]{Charles Guan}
\author[a]{Alexander P. Rockhill}
\author[b]{Masashi Sode}
\author[a,b]{Gianmarco Pinton}

\affil[a]{Forest Neurotech, Los Angeles, CA, USA}
\affil[b]{Lampe Joint Dept. of Biomedical Engineering, UNC-Chapel Hill and NC State University, Chapel Hill, NC, USA}

\cftpagenumbersoff{figure}
\cftpagenumbersoff{table}
\begin{document}
\maketitle

\begin{abstract}

\textbf{Purpose:}
  Volumetric ultrafast ultrasound produces massive datasets with high frame rates, dense reconstruction grids, and large channel counts.
  Beamforming computational demands limit research throughput and prevent real-time applications in emerging modalities such as elastography, functional neuroimaging, and microscopy.

\textbf{Approach:}
  We developed mach, an open-source, GPU-accelerated beamformer with a highly optimized delay-and-sum CUDA kernel and an accessible Python interface.
  mach uses a hybrid delay computation strategy that substantially reduces memory overhead compared to fully precomputed approaches.
  The CUDA implementation optimizes memory layout for coalesced access and reuses delay computations across frames via shared memory.
  We benchmarked mach on the PyMUST rotating disk dataset and validated numerical accuracy against existing open-source beamformers.

\textbf{Results:}
  mach processes 1.1 trillion points per second on a consumer-grade GPU, achieving $>$10$\times$ faster performance than existing open-source GPU beamformers.
  On the PyMUST rotating disk benchmark, mach completes reconstruction in 0.23~ms, 6$\times$ faster than the acoustic round-trip time to the imaging depth.
  Validation against other beamformers confirms numerical accuracy with errors below $-60$~dB for Power Doppler and $-120$~dB for B-mode.

\textbf{Conclusions:}
  mach achieves 1.1 trillion points per second throughput, enabling real-time 3D ultrafast ultrasound reconstruction for the first time on consumer-grade hardware.
  By eliminating the beamforming bottleneck, mach enables real-time applications such as 3D functional neuroimaging, intraoperative guidance, and ultrasound localization microscopy.
  mach is freely available at \url{https://github.com/Forest-Neurotech/mach}.

\end{abstract}

\keywords{3D beamforming, image reconstruction, ultrafast ultrasound imaging, open-source, GPU}

\section{Introduction}
\label{sect:intro}

Over the past two decades, biomedical ultrasound has expanded beyond 2D brightness-mode (B-mode) imaging into novel clinical applications.
Ultrafast ultrasound imaging~\cite{tanter2014-ultrafast} achieves frame rates up to 100 times faster than conventional focused methods, enabling researchers and clinicians to capture millisecond-scale physiology and quantify tissue characteristics.
Building on these advances, researchers have recorded myocardial stiffness~\cite{provost2014-3d}, neural activity~\cite{mace2011-fusi,montaldo2022-fusi}, and super-resolution vascular maps~\cite{errico2015-ulm, jones2024non}.

In parallel, novel transducer designs have extended ultrasound imaging from 2D to 3D~\cite{smith1991-3d}.
Conventional ultrasound acquires thin 2D slices and requires the operator to move the probe to build a mental model of the target anatomy.
Advanced 2D array transducers, such as matrix arrays, instead capture volumetric data in a single acquisition, simplifying the acquisition of 3D anatomy and physiology~\cite{sanchez2021-uoc}.
While matrix arrays remain primarily research tools~\cite{rabut2019-4d, jones2024non, heiles2025-sheet} rather than clinical standards, their 3D imaging capabilities open new possibilities when paired with ultrafast imaging.
Combining these technologies, 3D ultrafast ultrasound~\cite{provost2014-3d} has already enabled imaging of the entire rodent brain at unprecedented resolution~\cite{heiles2025-sheet} and functional sensitivity~\cite{rabut2019-4d, jones2024non}.

Despite these capabilities, 3D ultrafast imaging faces major computational barriers.
A typical acquisition samples hundreds of elements at multi-megahertz rates, generating 1~gigabyte per second, or terabytes per session.
Beamforming, the process of reconstructing images from raw sensor data, requires computational throughput that existing frameworks cannot achieve in real time (Sec.~\ref{sect:technical-background}).
This bottleneck slows research~\cite{jones2024non} and limits real-time feedback~\cite{griggs2024-bci}, which would be valuable for guided interventions.

Here, we introduce mach, an open-source, GPU-accelerated Python beamforming library that overcomes these computational barriers.
mach beamforms more than 10$\times$ faster than existing open-source GPU beamformers (Sec.~\ref{sect:performance}), enabling researchers and clinicians to reconstruct 3D ultrasound volumes as they are acquired.
Beyond raw speed, mach emphasizes accessibility: it installs as a standard Python package, integrates with scientific computing tools, includes comprehensive documentation and reproducible examples, and runs on widely available consumer GPUs (Sec.~\ref{sect:usage}).
By eliminating the beamforming bottleneck, mach makes advanced volumetric imaging a practical tool for the ultrasound research and clinical communities.

\section{Technical Background}
\label{sect:technical-background}

\subsection{Delay-and-sum beamforming}
\label{subsect:das-background}

\begin{figure}[ht]
    \centering
    \includegraphics[width=0.85\linewidth]{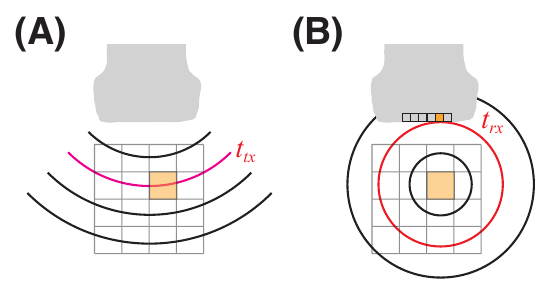}
    \caption{Transmit and receive geometry for pulse-echo ultrasound imaging. (a) Transmitted wavefront at different times, with a highlighted voxel corresponding to arrival time $t_{tx}$. In ultrafast imaging, the wavefront is typically unfocused and insonifies the entire field of view. (b) Point scatterers in the medium re-emit spherical waves that are received by all array elements. The return path time at the highlighted transducer element from the same voxel in (a) is $t_{rx}$.}
    \label{fig:transmit_receive_probe}
\end{figure}

In pulse-echo ultrasound imaging, a transducer insonifies the target with an acoustic pulse and receives the backscattered echo signals (Fig.~\ref{fig:transmit_receive_probe}).
These signals are then beamformed to reconstruct an image.
The most widespread beamforming technique is delay-and-sum (DAS)~\cite{perrot2021-das}, favored for its efficiency and robustness.

DAS beamforming reconstructs images using simplified wave propagation physics.
The algorithm proceeds as follows: (1)~for each voxel, compute the round-trip acoustic propagation \emph{delays} from the transmitting transducer to the voxel (Fig.~\ref{fig:transmit_receive_probe}a) and back to \emph{each} receiving element (Fig.~\ref{fig:transmit_receive_probe}b); (2)~index or interpolate the received signals at those delay times; (3)~sum the delayed signals coherently across all receiving elements.
The propagation delays depend on the transmitted wavefront (focused, plane-wave, or diverging), the positions of the receiving elements, and the speed of sound in the medium (typically 1540~m/s in soft tissue).
When the delays are correct, they align the phases of the received signals across elements, such that backscattered echoes from the same voxel add constructively, while noise adds incoherently and tends to cancel~\cite{perrot2021-das}.
The resulting summed signal represents the value of that voxel.

The delay-and-sum algorithm repeats for every voxel in the volume and every acquisition frame.
Many acquisition sequences also \emph{compound} multiple transmit angles or virtual source positions to improve image quality~\cite{rabut2019-4d, jones2024non}, further increasing the computational load.
The number of delays to calculate, or equivalently, the number of points to sum, can be expressed as~\cite{vbeam}:

\begin{equation}
  \text{Throughput (pts/s)} = \text{Voxels} \times \text{Channels} \times \text{Compounding} \times \text{Frame rate}
  \label{eq:throughput}
\end{equation}

In other words, the computational cost of DAS scales with the reconstruction grid size, transducer channel count, and the transmit-receive rate.
As a result, the computational demands of beamforming scale dramatically from conventional 2D imaging to 3D ultrafast imaging (Table~\ref{tab:throughput-requirements}).
Ultrafast 3D imaging can require throughputs exceeding one trillion points per second, around 1000$\times$ greater than that of conventional 2D imaging.

\begin{table}[htbp]
\caption{Typical beamforming throughput requirements across imaging modalities.}
\begin{center}
\small
\begin{tabular}{l|ccc|c}
\toprule
\textbf{Imaging mode} & \textbf{Voxels}$^a$ & \textbf{Channels} & \textbf{Frame rate}$^b$ & \textbf{Required throughput}$^c$ \\
 & & & \textbf{(Hz)} & \textbf{($10^{9}$ pts/s)} \\
\midrule
2D+t Conventional \cite{schmitz2020ultrasoundimaging} & 192 $\times$ 512 & 128 & 78 & 1 \\
2D+t Ultrafast \cite{griggs2024-bci} & 128 $\times$ 160 & 128 & 11 $\times$ 500 & 14 \\
3D+t Ultrafast \cite{rabut2019-4d} & 64 $\times$ 64 $\times$ 175 & 512 & 8 $\times$ 390 & 1145 \\
\bottomrule
\end{tabular}
\label{tab:throughput-requirements}
\end{center}
\vspace{1ex}
\noindent \textit{Note:} Representative examples shown; parameters vary by study. \\[0.5ex]
$^a$Width $\times$ height (2D+t) or width $\times$ length $\times$ height (3D+t) voxel dimensions. \\
$^b$Frame rate includes the compounding factor for ultrafast methods. \\
$^c$Throughput calculated as in Equation~\ref{eq:throughput}.
\end{table}

\subsection{Open-source beamformers}
\label{subsect:open-source}

The ultrasound research community has developed several open-source beamforming platforms.
Table~\ref{tab:beamformer-comparison} summarizes recent open-source beamformers~\cite{ultraspy,vbeam,rtbf,pymust,must,ustb}, their GPU support, and their throughputs.
Many are well-documented, highly flexible, and achieve throughputs of tens of billions of points per second~\cite{ultraspy,vbeam,rtbf}, well-suited for 2D ultrafast imaging.
However, as shown in Table~\ref{tab:throughput-requirements}, 3D ultrafast imaging requires much higher throughputs than existing beamformers achieve.

\begin{table}[htbp]
\caption{Recent open-source ultrasound beamformers.}
\begin{center}
\small
\begin{tabular}{lllll}
\toprule
\textbf{Library} & \textbf{Language} & \textbf{GPU} & \textbf{Key Features} & \textbf{Throughput}$^a$ \\
 & & & & \textbf{($10^{9}$ pts/s)} \\
\midrule
(Py)MUST \cite{pymust, must} & Python, MATLAB & No & simulation tools & --- \\
USTB \cite{ustb} & MATLAB/C++ & Optional & general beamformer & --- \\
ultraspy \cite{ultraspy} & Python/CUDA & Optional & additional algorithms & 34 \\
vbeam \cite{vbeam} & Python (JAX) & Optional & differentiable & 64 \\
rtbf \cite{rtbf} & MATLAB/CUDA & Required & real-time & 83 \\
mach (this work) & Python/CUDA & Required & 3D ultrafast beamforming & 1130 \\
\bottomrule
\end{tabular}
\label{tab:beamformer-comparison}
\end{center}
\vspace{1ex}
\noindent $^a$Delay-and-sum throughput values were estimated using Equation~\ref{eq:throughput} from published results where available. GPU hardware specifications vary across publications. A dash (---) indicates throughput was not reported. mach throughput is described in Sec.~\ref{sect:performance}.
\end{table}

\subsection{GPU programming}
\label{subsect:cuda}

As described in Sec.~\ref{subsect:das-background}, delay-and-sum beamforming repeats independently across voxels and frames.
Like graphics rendering, this structure is well-suited for GPU acceleration.
Achieving real-time throughput for 3D ultrafast imaging requires careful attention to GPU architecture and memory organization.
This section provides a brief overview of GPU programming concepts relevant to beamforming optimization, focusing on the CUDA programming model~\cite{cudaprog}.

CUDA applications start on the CPU to configure and launch functions, also called \emph{kernels}, that execute on the GPU.
A kernel launch creates many---often millions of---parallel threads, each executing the same GPU code on different data.
GPU kernels achieve high throughput when structured so that different threads require minimal synchronization.
In beamforming frameworks, threads often contribute to one or more delay calculations or voxels, parallelizing the computation across the reconstruction grid.

GPU threads are organized hierarchically into warps (32 threads) and thread blocks, which execute on streaming multiprocessors (SMs).
All threads within a thread block execute on the same SM, enabling efficient synchronization between threads.
Different thread blocks may execute in any order on any available SM, allowing the same kernel to scale across GPUs with varying SM counts, from mobile devices to data centers.

Modern GPUs contain multiple memory spaces with distinct performance characteristics:
\begin{itemize}
\item Global DRAM: Large but slow. As an example, the GeForce RTX 5090 has 32~GB with peak bandwidth of $\sim$1.8~TB/s.
\item L2 cache: Smaller but faster. The RTX 5090 has 98~MB of L2 cache.
\item Shared memory / L1 cache: Very fast but limited. Each SM on the RTX 5090 has 128~KB of combined L1 cache and shared memory, with latency on the order of clock cycles. For beamforming, delay computations can be cached here and reused across frames (Sec.~\ref{subsect:algorithmic-optimizations}).
\end{itemize}
Effective GPU programming keeps frequently accessed data in fast memory tiers and minimizes global memory access.

Additionally, global memory is accessed in 32-byte transactions.
When threads in the same warp access consecutive memory addresses, the GPU \emph{coalesces} these requests into a single transaction.
Conversely, when threads access scattered addresses, each may require its own transaction.
A seemingly simple change, transposing data layout to favor coalesced access, can improve effective memory throughput by $\sim$8$\times$ \cite{cudaprog}.
Because global memory bandwidth is often limiting, coalesced access can substantially improve kernel performance (Sec.~\ref{subsect:cuda-optimizations}).

\section{Software usage}
\label{sect:usage}

mach is a GPU-accelerated beamforming library designed for real-time ultrafast ultrasound imaging.
It provides a Python module for delay-and-sum beamforming, allowing users to integrate their own acquisition interfaces and postprocessing pipelines.
The documentation includes end-to-end workflow examples, and the source code is freely available under the BSD-3-Clause-Clear license at \url{https://github.com/Forest-Neurotech/mach}.

\subsection{Installation}

mach is available on the Python Package Index (PyPI) and can be installed with pip:

\begin{verbatim}
pip install mach-beamform
\end{verbatim}

In the current (as of writing) version of mach (0.1.0), pip installation requires Linux, CUDA 12.3 or newer, and a CUDA-compatible GPU.
For users without local GPU access, mach can run on cloud platforms such as Google Colab.

Installation from source (\url{https://github.com/Forest-Neurotech/mach}) is an alternative for contributors or users on non-Linux systems.
A Docker image is also available on the GitHub Container Registry.

\subsection{Documentation}
\begin{figure}[htbp]
\centering
\includegraphics[width=0.9\textwidth]{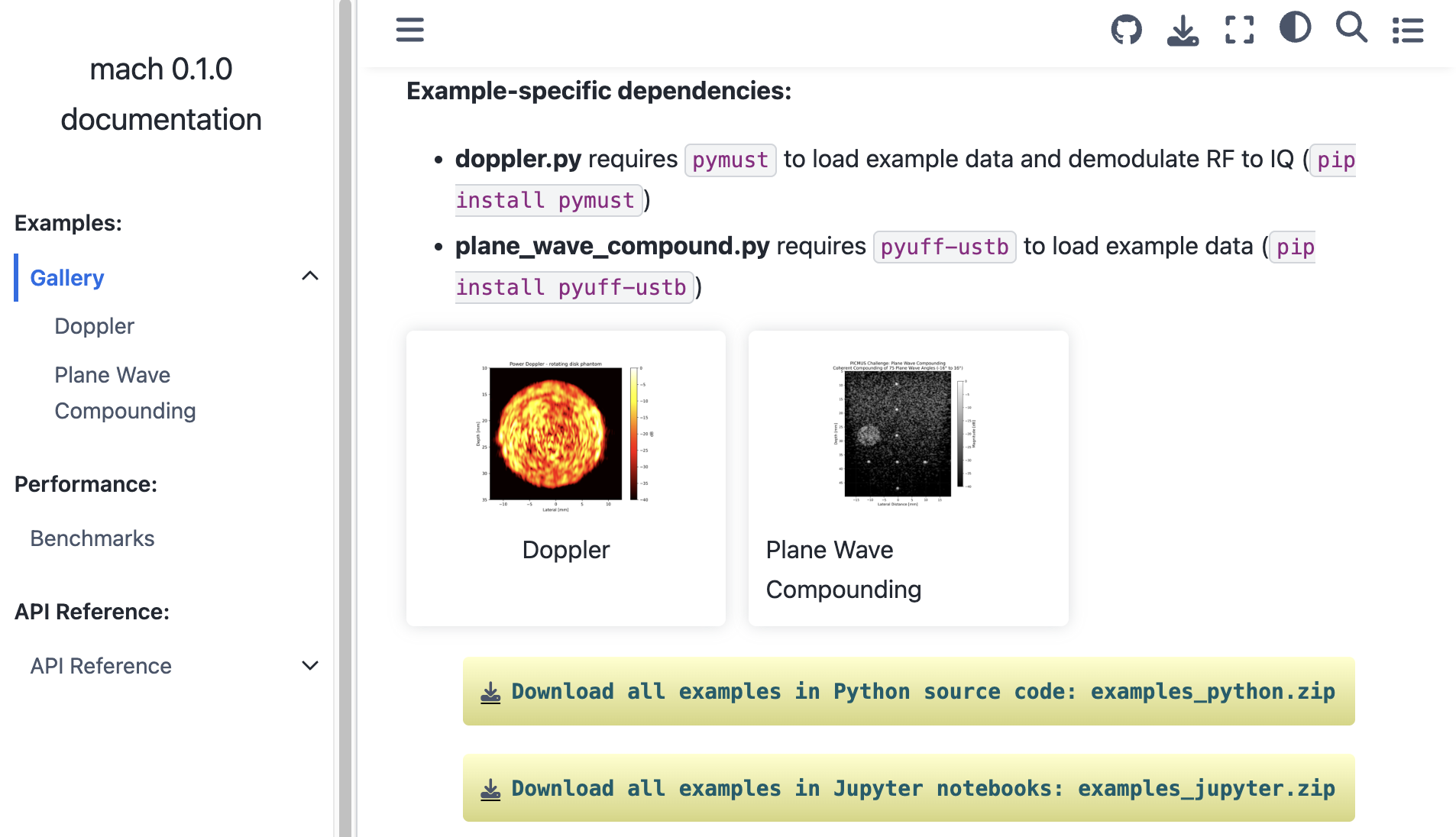}
\caption{Screenshot of the mach documentation website showing the example gallery.}
\label{fig:example-gallery}
\end{figure}

Documentation is available at \url{https://forest-neurotech.github.io/mach/} (Fig.~\ref{fig:example-gallery}) and includes installation instructions, interactive Jupyter notebook examples, and the application programming interface (API) reference.

\subsection{Example workflow}

mach takes as input raw ultrasound data (RF or IQ), acquisition parameters such as the transmitted wavefronts, and metadata including transducer geometry, center frequency, sampling rate, and the assumed speed of sound in the medium.

A typical workflow consists of:

\begin{enumerate}
\item Load raw ultrasound data, sequence parameters, and metadata from a recording.
\item Define the reconstruction grid.
\item Precompute transmit wavefront arrival times at the reconstruction grid, e.g., for each plane-wave angle (Sec.~\ref{subsect:algorithmic-optimizations}). External simulation tools~\cite{fullwave25,must} can also compute arrival times for complex wavefronts or non-homogeneous media.
\item Call the mach beamforming function to reconstruct volumes from raw ultrasound frames.
\end{enumerate}

Additional postprocessing steps can follow, such as log-compression for B-mode imaging or clutter filtering for Doppler imaging~\cite{demene2015}.

\begin{figure}[htbp]
\centering
\includegraphics[width=0.9\textwidth]{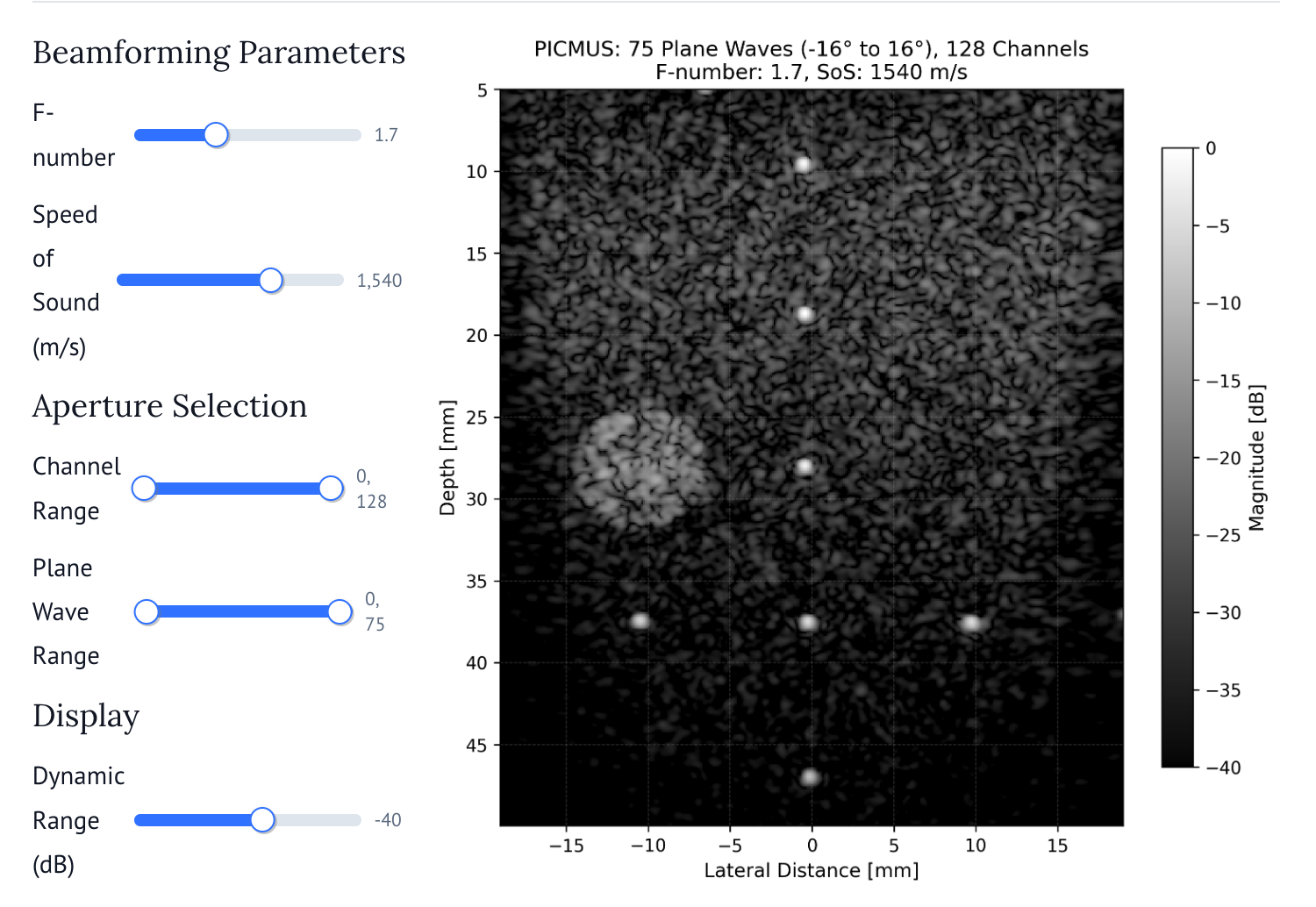}
\caption{Example, interactive marimo user-interface to modify mach beamforming parameters on a PICMUS dataset.}
\label{fig:marimo-notebooks}
\end{figure}

The documentation's example gallery (Fig.~\ref{fig:example-gallery}) demonstrates this workflow and is the recommended entry point for adapting mach to new datasets.
Interactive examples are also available as marimo~\cite{marimo} notebooks in the source repository (Fig.~\ref{fig:marimo-notebooks}).

\subsection{Compatibility with scientific Python ecosystem}

Through the Python Array API standard~\cite{array_api}, mach works directly with popular array libraries including NumPy~\cite{numpy}, PyTorch~\cite{pytorch}, JAX~\cite{jax}, and CuPy~\cite{cupy}.
This design allows researchers to integrate mach into existing workflows.
Compatibility also improves efficiency: GPU-backed input arrays remain on the GPU, avoiding unnecessary CPU--GPU transfers.
Beamformed results can likewise remain on the GPU for downstream processing such as clutter filtering or machine learning inference.

\section{Performance}
\label{sect:performance}

As shown in Table~\ref{tab:throughput-requirements}, 3D ultrafast imaging requires much faster throughputs than conventional 2D imaging.
This section quantifies mach's performance using standard benchmarks and compares the results against existing open-source tools.

\subsection{Methods}

We evaluated mach using the publicly available PyMUST rotating-disk Doppler dataset~\cite{pymust}, a common~\cite{ultraspy} example dataset for ultrafast imaging.
This dataset provides a realistic 2D ultrafast workload while remaining tractable across beamformer implementations and hardware configurations.
It comprises 32 ultrafast plane-wave frames acquired with a 128-element linear array (L7-4), imaging a rotating disk target at depths from 10~mm to 35~mm.

The benchmark uses the following parameters:
\begin{itemize}
\item \textbf{Acquisition:} 32 ultrafast plane-wave frames, 128-element linear array (L7-4)
\item \textbf{Reconstruction grid:} $251 \times 251 = 63,001$ pixels, 0.1~mm spacing
\item \textbf{Imaging depth:} 35~mm maximum depth
\item \textbf{Data format:} complex64
\item \textbf{Beamforming parameters:} linear interpolation, f-number = 1.0
\end{itemize}

This configuration totals $251 \times 251 \times 128 \times 32 = 258$ million delay-and-sum operations (Equation~\ref{eq:throughput}).

We used the following hardware and software environment:
\begin{itemize}
\item \textbf{GPU:} NVIDIA GeForce RTX 5090 (32~GB VRAM, CUDA 12.8)
\item \textbf{CPU:} Intel Core Ultra 9 285K (64~GB RAM)
\item \textbf{Operating system:} Linux 6.11.0
\item \textbf{Python version:} 3.11
\item \textbf{mach version:} 0.1.0
\end{itemize}

Our timing measurements isolate the core delay-and-sum kernel, the most computationally intensive part of beamforming and thus the most important to optimize.
We exclude data transfer between CPU and GPU memory, which depends on PCIe configuration and whether the pipeline preprocesses or postprocesses on the GPU.
We also exclude preprocessing steps such as IQ demodulation or file-format conversion, which vary by scanner manufacturer.

For comparison, we ran the same benchmarks on vbeam (GPU-accelerated via JAX)~\cite{vbeam} and PyMUST (CPU implementation)~\cite{pymust} using the same hardware.
Both tools used their default settings and optimizations.

All benchmarks can be reproduced using the test suite included with mach's source code (\url{https://github.com/Forest-Neurotech/mach}).

\begin{figure}[htbp]
\centerline{\includegraphics[width=0.7\columnwidth]{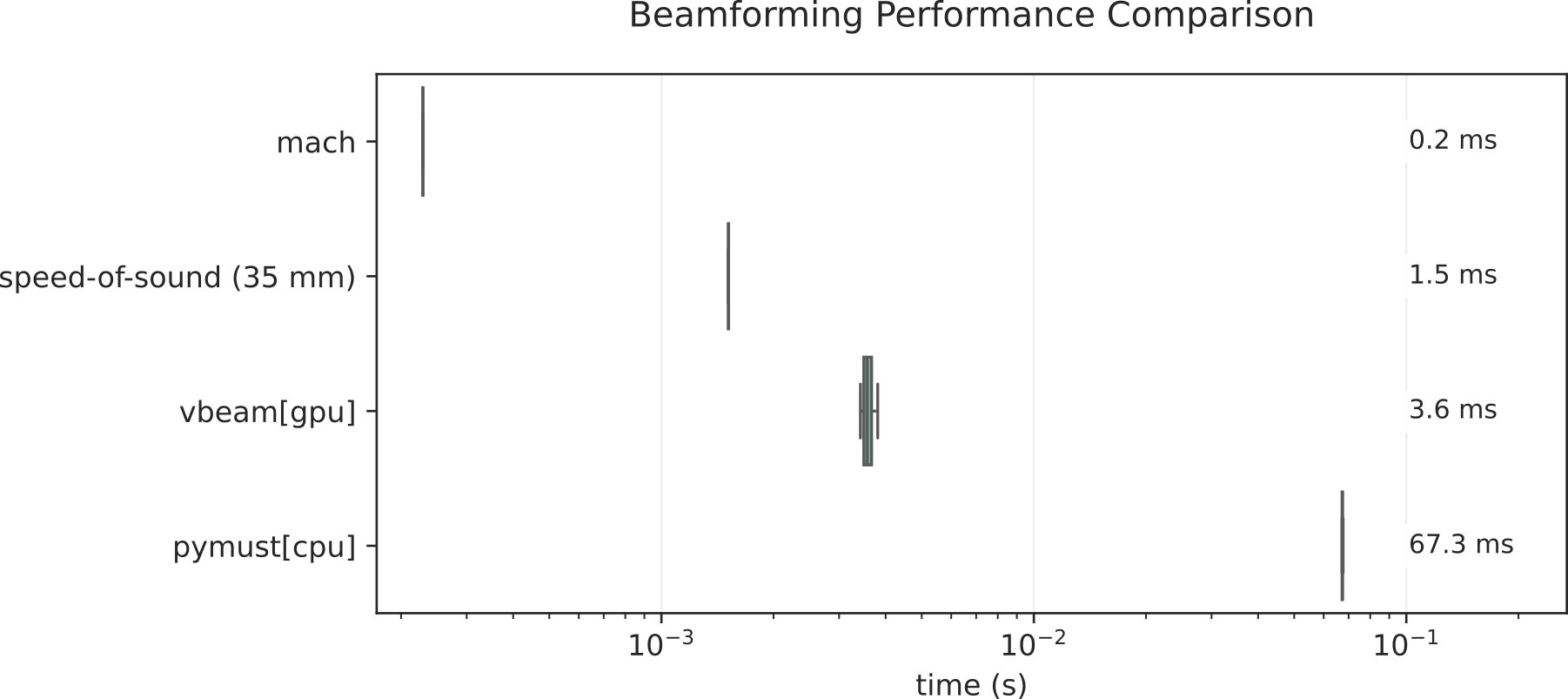}}
\caption{Runtime comparison. mach beamforms the rotating-disk dataset in 0.23~ms (1.1~Tpoints/s), 6$\times$ faster than the speed of sound.}
\label{fig:benchmark}
\end{figure}

\subsection{Results}
\label{subsect:performance-results}

mach beamforms the dataset in 0.23~ms (Fig.~\ref{fig:benchmark}), corresponding to 1.1 trillion points per second.
For context, this reconstruction time is 6$\times$ faster than the acoustic round-trip time to the maximum imaging depth: $(35\text{ mm} \times 2) / (1480\text{ m/s}) \times 32\text{ frames} \approx 1.5\text{ ms}$.
At this speed, mach could reconstruct 2D ultrafast acquisitions at frame rates exceeding 50~kHz, well beyond the Nyquist requirements of most applications.

The performance advantage is substantial: mach is $>$15$\times$ faster than vbeam (3.6~ms) and $>$290$\times$ faster than PyMUST's CPU implementation (67.3~ms).
These comparisons should not diminish the contributions of PyMUST and vbeam, which have been instrumental to ultrasound research and provided important foundations for mach's development.
Rather, they illustrate mach's optimization focus: accelerating high-channel-count, ultrafast imaging workflows.
The following sections explain how these gains are achieved.

\subsection{Algorithmic optimizations}
\label{subsect:algorithmic-optimizations}

The first key to mach's performance is a strategic design decision about \emph{what} to compute and what to precompute.

Delay-and-sum beamforming requires computing the round-trip acoustic path from each transmit wavefront, through each voxel, to each receive element~\cite{perrot2021-das}.
Although input channel data changes per frame, the delays remain constant.
This creates a memory-computation trade-off: some beamformers precompute and cache all delays to minimize per-frame computation~\cite{perrot2021-das,pymust,must}, but fully precomputed delays require prohibitive memory for 3D volumes with high-channel-count arrays~\cite{perrot2021-das}.
Conversely, other beamformers compute all delays on the fly~\cite{vbeam, ultraspy}, which is expensive due to the square roots and divisions involved.

mach adopts a hybrid approach that balances memory and computation.
The arrival times of each transmit wavefront at each voxel are precomputed and stored once, a modest memory requirement in typical ultrafast acquisitions (Table~\ref{tab:memory-calculation}).
For additional flexibility, external simulation tools~\cite{fullwave25,must} can compute arrival times for complex wavefronts or non-homogeneous media.
Receive delays are computed on the fly within each CUDA thread block.
This strategy achieves substantial memory savings compared to fully precomputed delays while keeping per-frame computation lightweight.

\subsection{CUDA optimizations}
\label{subsect:cuda-optimizations}

The second key to mach's performance is how the algorithm is implemented on the GPU.
These optimizations target the specific characteristics of delay-and-sum to maximize memory bandwidth and minimize redundant computation.

First, mach organizes channel data with the frame dimension stored contiguously (shape: [elements, samples, frames]) to maximize memory coalescing.
This layout ensures threads within a warp access consecutive addresses when processing temporal ensembles, allowing the GPU to coalesce requests into high-bandwidth transactions.

Second, mach extends the delay-reuse strategy (Sec.~\ref{subsect:algorithmic-optimizations}) to the thread block level.
In addition to reusing transmit wavefront arrival times across kernel invocations, mach reuses delay calculations across frames within a thread block (e.g., 32 frames of 8 voxels).
Receive delays are computed once per voxel and cached in shared memory for reuse across all temporal frames.
Since delay calculations are expensive (involving square roots and divisions), this reuse eliminates redundant per-frame computation.

These optimizations collectively achieve 90\% memory throughput and 69\% compute utilization (Table~\ref{tab:nsight}), approaching the practical performance limits where further optimization yields diminishing returns.

\begin{table}[htbp]
\caption{GPU Compute and Memory Utilization, measured via NVIDIA Nsight Compute}
\begin{center}
\begin{tabular}{lc}
\toprule
\textbf{Resource} & \textbf{Utilization} \\
\midrule
Compute& 69\%\\
Memory (Overall)  & 90\%\\
Memory (L1 Cache)          & 45\%\\
Memory (L2 Cache)          & 90\%\\
Memory (DRAM)              & 8\%\\
\bottomrule
\end{tabular}
\label{tab:nsight}
\end{center}
\end{table}

\subsection{Scaling}
\label{subsect:scaling}

While the PyMUST benchmark represents a typical 2D ultrafast acquisition, 3D ultrafast imaging with high-channel-count arrays poses greater computational demands.
Table~\ref{tab:throughput-requirements} includes a representative 3D+t functional ultrasound experiment that processes ensembles of 136 compounded volumes into a single Doppler volume, requiring nearly 400 billion delay-and-sum operations.

\begin{figure}[htbp]
\centerline{\includegraphics[width=\columnwidth]{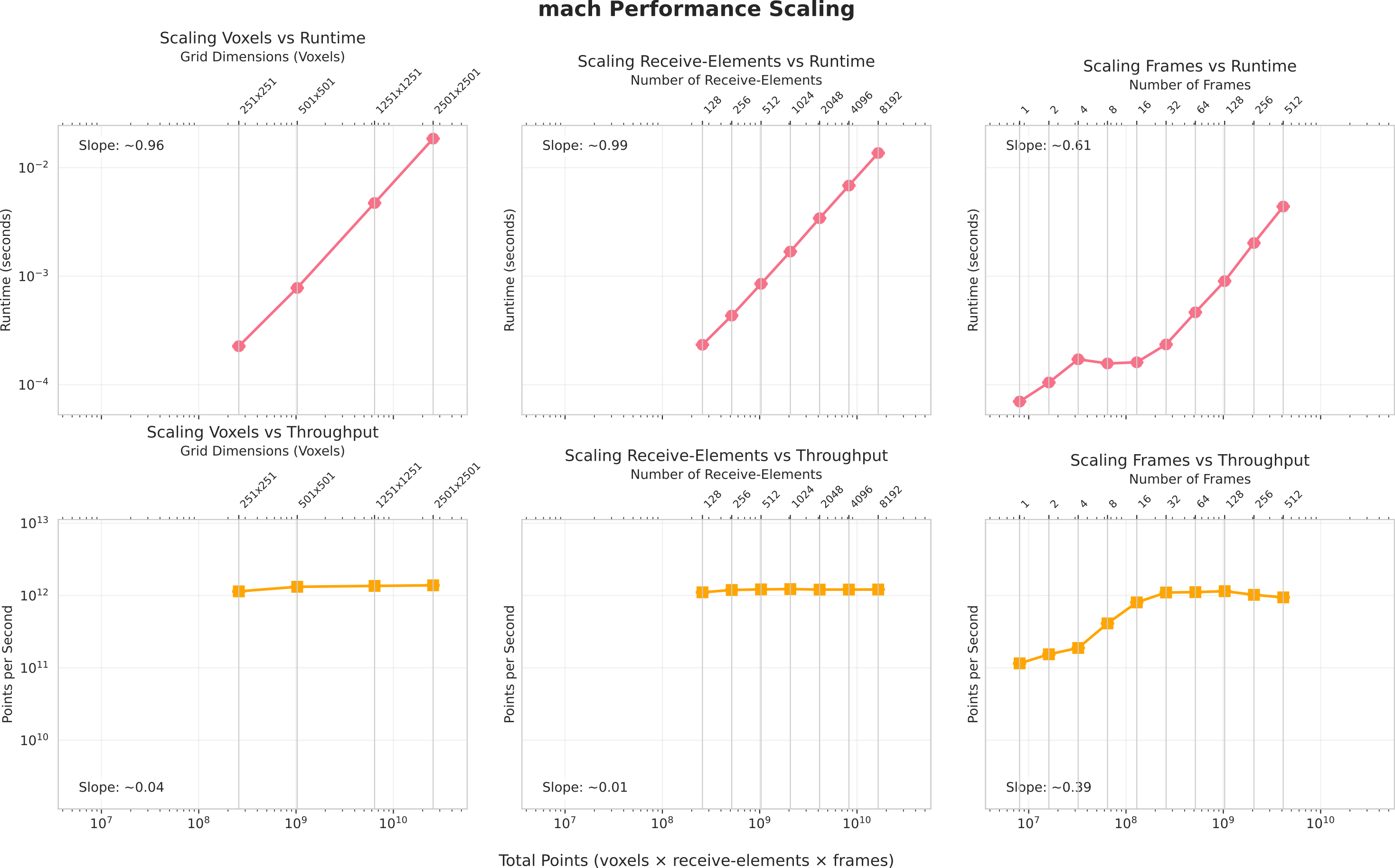}}
\caption{Performance scaling of mach. We independently varied the number of voxels, elements, and frames in the PyMUST rotating disk dataset and measured runtime (top row) and throughput (bottom row). Runtime scales near-linearly with the number of voxels (left) and elements (middle), resulting in constant throughput. Throughput (bottom right) decreases for frame counts below 16 as memory access becomes less coalesced.}
\label{fig:scaling}
\end{figure}

To evaluate scaling, we extended the benchmarks by varying voxel count (63k to 6.3M), channel count (128 to 8,192), and ensemble size (1 to 512 frames).
Figure~\ref{fig:scaling} shows that mach maintains consistent per-point throughput ($>$1 trillion points/second) across these ranges.
Minor performance variations occur only for very small ensemble sizes ($<$16 frames for complex64 data), where insufficient temporal samples limit coalescing efficiency.
For typical research workloads ($\geq$32 frames), mach achieves the throughput needed for real-time 3D ultrafast imaging (Table~\ref{tab:throughput-requirements}).

mach's GPU memory usage scales as:

\begin{equation}
  \text{Memory} \propto (\text{Voxels} \times \text{Frames}) + (\text{Channels} \times \text{Samples} \times \text{Frames})
  \label{eq:memory}
\end{equation}

Table~\ref{tab:memory-calculation} shows the memory calculation for the representative 3D+t fUSI acquisition in Table~\ref{tab:throughput-requirements}, using 8 compound planes $\times$ 136 frames = 1,088 temporal frames and 384 samples per channel~\cite{rabut2019-4d}.
Total GPU memory usage is approximately 8~GB, well within the 16--32~GB capacity of consumer GPUs.
This makes mach accessible to researchers using widely available hardware rather than specialized data-center GPUs.

\begin{table}[htbp]
\caption{GPU memory usage for a typical 3D+t ultrafast fUSI acquisition \cite{rabut2019-4d}}
\label{tab:memory-calculation}
\centering
\small
\begin{tabular}{lcccc}
\toprule
\textbf{Array} & \textbf{Dimensions} & \textbf{Data type} & \textbf{Bytes per} & \textbf{Memory} \\
 & & & \textbf{element} & \textbf{(GB)} \\
\midrule
Acquisition channel data & $512 \times 384 \times 1{,}088$ & complex64 & 8 & 1.7 \\
Receive-element coordinates & $512 \times 3$ & float32 & 4 & $<$0.01 \\
Output-grid coordinates & $716{,}800 \times 3$ & float32 & 4 & $\sim$0.01 \\
Wavefront arrival times & $716{,}800$ & float32 & 4 & $<$0.01 \\
Output beamformed volumes & $716{,}800 \times 8 \times 136$ & complex64 & 8 & 6.2 \\
\midrule
\textbf{Total} & & & & \textbf{$\sim$8~GB} \\
\bottomrule
\end{tabular}
\end{table}

\section{Validation}
\label{sect:validation}

To validate numerical correctness, we compared mach against established open-source beamformers PyMUST~\cite{pymust,perrot2021-das} and vbeam~\cite{vbeam}.

\begin{figure}[htbp]
\centerline{\includegraphics[width=\columnwidth]{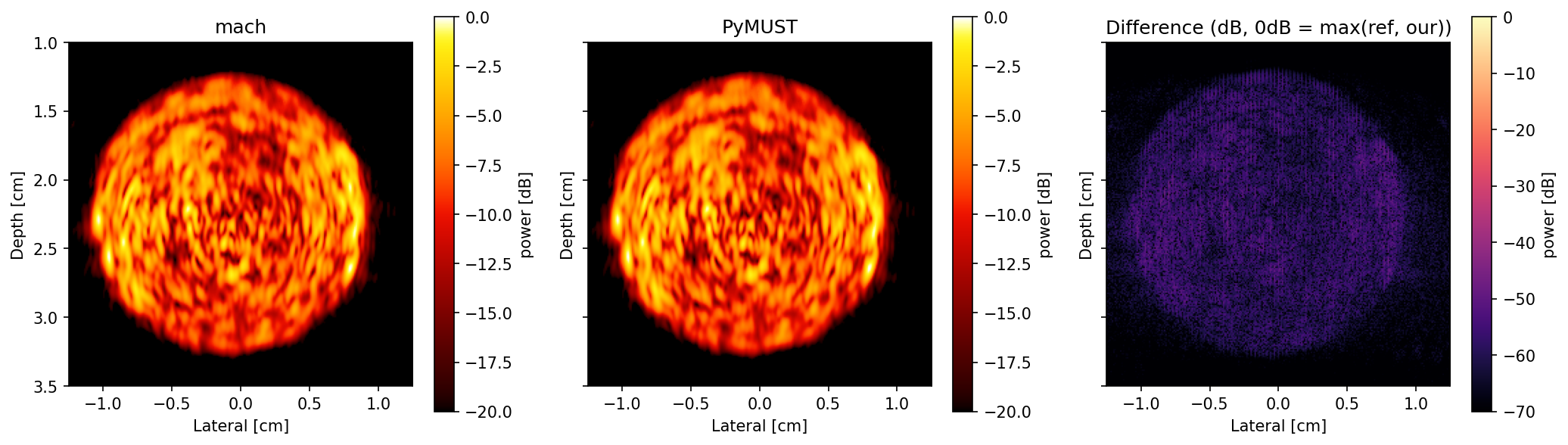}}
\caption{Validation of mach against PyMUST on the rotating disk Doppler dataset. Power Doppler images generated by mach (left) and PyMUST (center) match visually. The difference image (right) shows negligible error (below $-60$~dB of peak power), validating mach's numerical accuracy.}
\label{fig:validation_pymust}
\end{figure}

\begin{figure}[htbp]
\centerline{\includegraphics[width=\columnwidth]{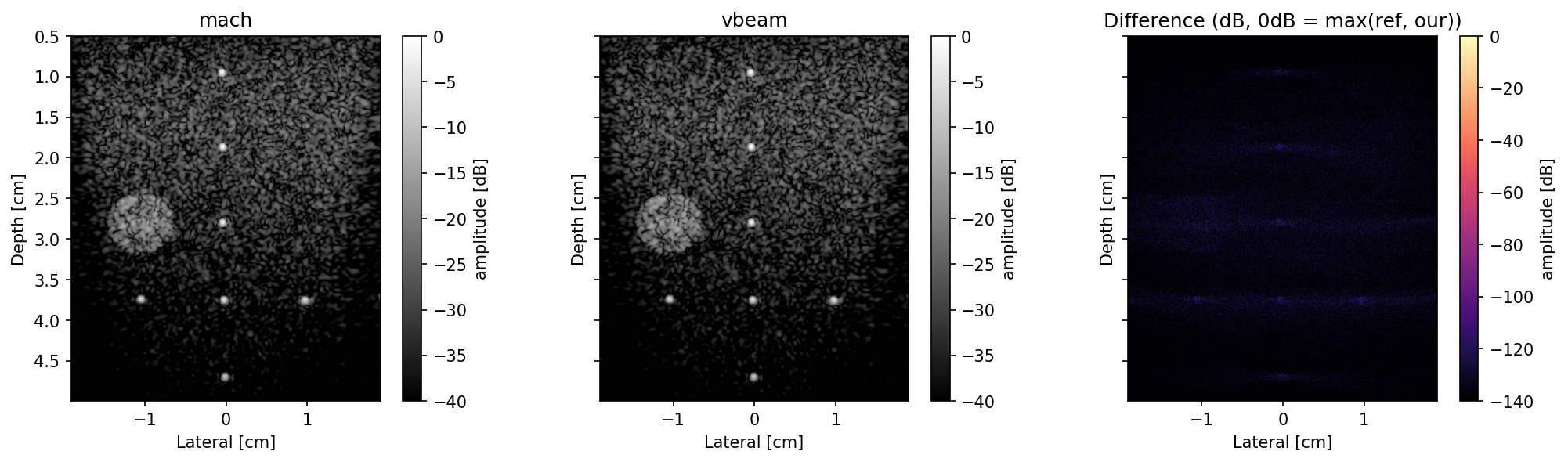}}
\caption{Validation of mach against vbeam on the PICMUS resolution and distortion dataset. B-mode images of point and cyst phantom targets generated by mach (left) and vbeam (center). The difference image (right) shows negligible error (below -120~dB of the peak value), validating mach's numerical accuracy.}
\label{fig:validation_vbeam}
\end{figure}

We used mach and PyMUST to beamform the PyMUST rotating disk dataset and formed a Power Doppler image.
Figure~\ref{fig:validation_pymust} shows that mach and PyMUST produce visually identical results, with pixel errors below $-60$~dB (power; 0~dB is the peak value in the reference image).
Next, we validated mach against vbeam on the PICMUS resolution/distortion dataset~\cite{picmus}.
Figure~\ref{fig:validation_vbeam} shows that the resulting B-mode images are visually identical, with pixel errors below $-120$~dB (amplitude).
These results confirm that mach is numerically equivalent to existing validated implementations.

\subsection{Continuous integration testing}

All benchmarks, validation tests, and documentation builds run automatically via GitHub Actions on every commit to the main branch, ensuring the latest release is accurate and reproducible.
Users can run the test suite locally to verify installation and numerical accuracy.

\section{Application: 3D ultrasound localization microscopy}
\label{sect:application-ulm}

As a practical example, we used mach to beamform a 3D ultrasound localization microscopy (ULM) dataset.
ULM~\cite{errico2015-ulm} creates super-resolution images of microvasculature by tracking individual microbubbles flowing through the circulatory system.
Over time, these microbubbles map out the vasculature, such that the 3D+t recording can be aggregated into a single 3D volume, much like a long-exposure nighttime photograph.
To precisely track microbubble movement, ULM requires high-resolution reconstruction grids, sustained recordings ($>$1~minute), and ultrafast frame rates ($>$500~Hz) to spatiotemporally filter microbubble motion from background noise.
Extending ULM to 3D+t previously required large-scale, offline beamforming~\cite{jones2024non}.

We used mach to beamform a 3D+t ULM rat dataset acquired transcranially with a 32$\times$32 Vermon matrix probe transmitting at 7.81~MHz~\cite{jones2024non}.
5 compounding angles/volume $\times$ 500 volumes/second were acquired for 200 seconds.
The reconstruction grid was 94 $\times$ 94 $\times$ 178 voxels with isotropic 96~$\mu$m resolution.
We applied a volumetric ULM postprocessing pipeline~\cite{jones2024non} to convert the 5 $\times$ 100,000 beamformed volumes into a 3D rendering of the microvasculature.

\begin{figure}[htbp]
\centering
\includegraphics[width=0.3\textwidth]{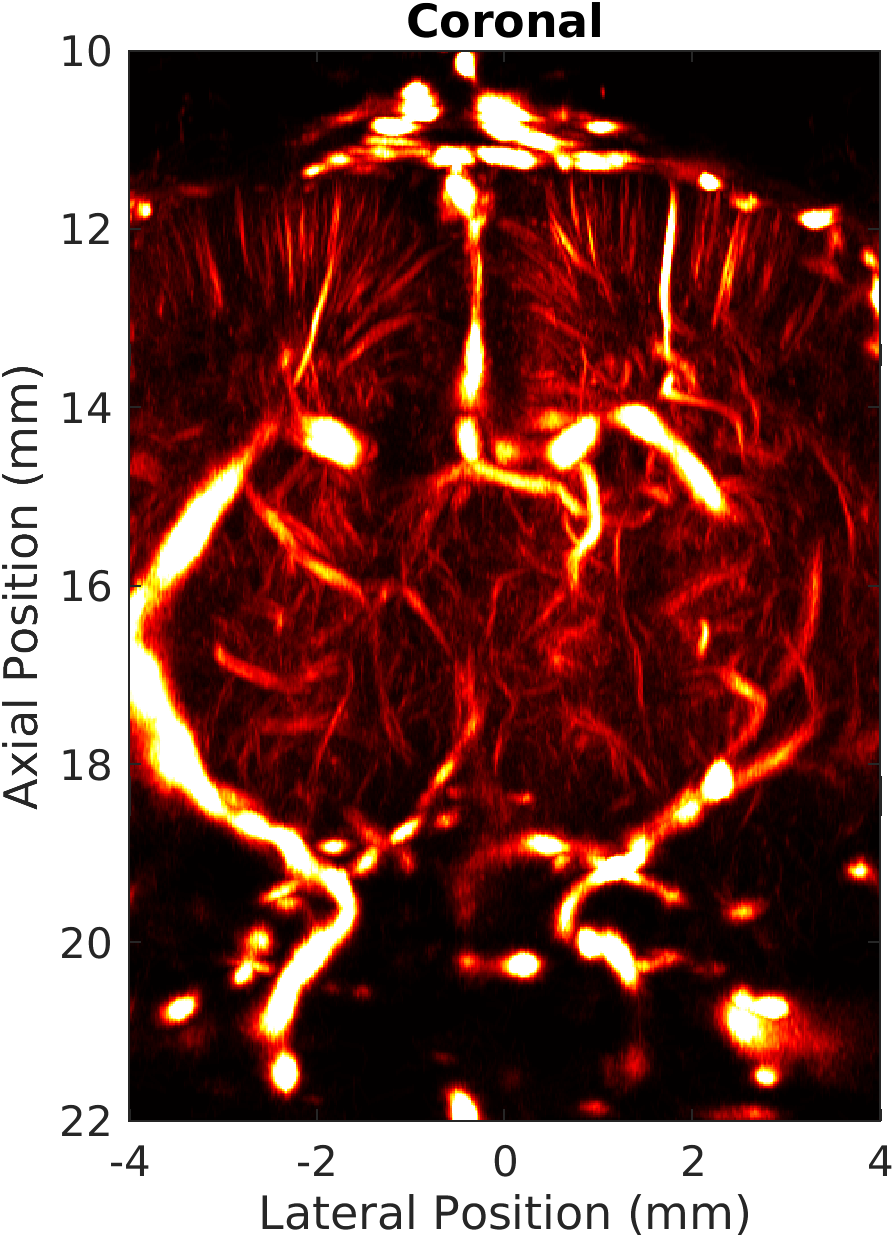}
\includegraphics[width=0.3\textwidth]{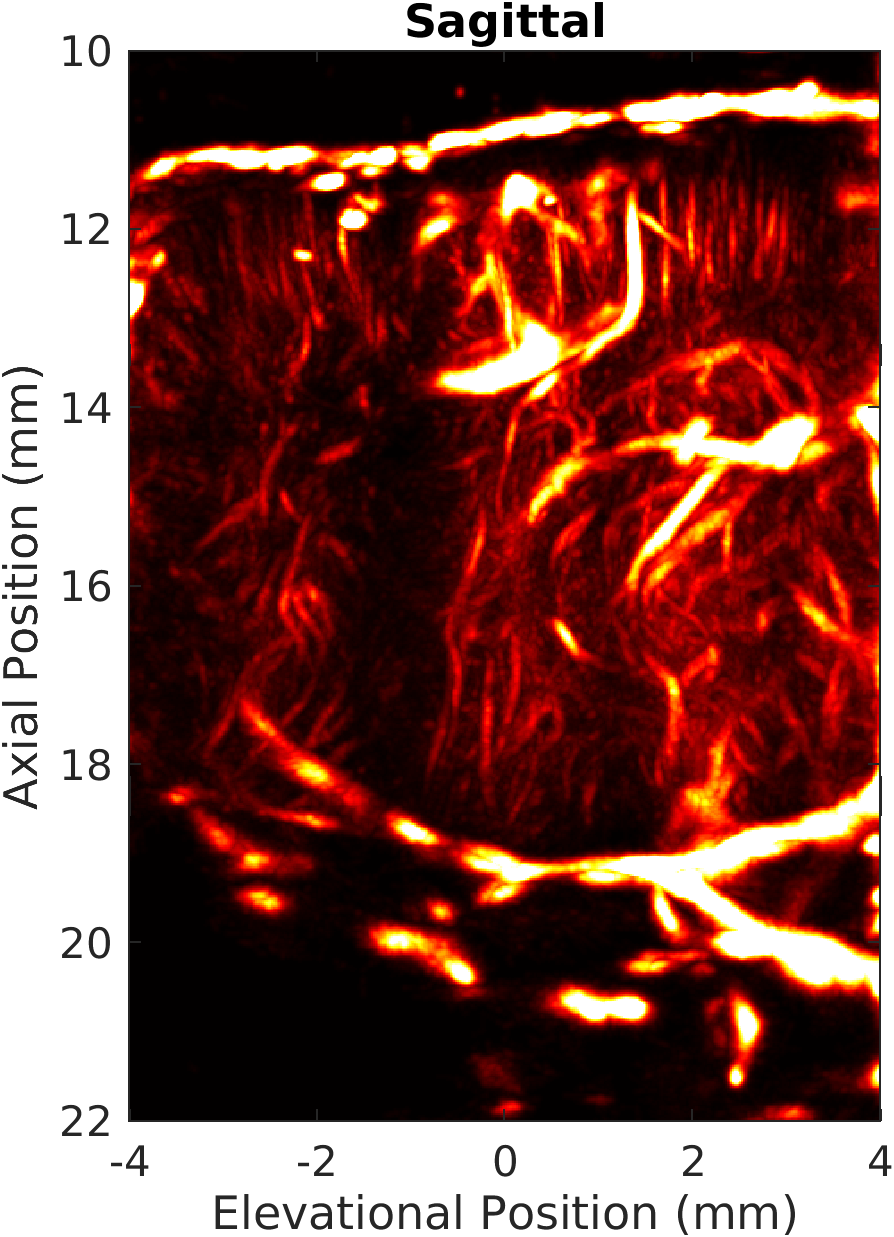}
\caption{3D ultrasound localization microscopy of rat brain microvasculature. Maximum-intensity projections along the elevational (left) and lateral (right) axes show super-resolved vascular structures.}
\label{fig:ulm}
\end{figure}

Figure~\ref{fig:ulm} shows a maximum-intensity projection of the super-resolved rat brain microvasculature.
Despite attenuation from the skull, the recording captured fine vascular structures.

mach's high throughput enabled processing the entire 200-second acquisition (100,000 volumes; 500,000 before compounding) in a reasonable time frame.

We profiled the delay-and-sum kernel, the most compute-intensive part of the pipeline.
mach beamformed 1,572,808 voxels $\times$ 1024 channels $\times$ 5 angles $\times$ 124 frames/second $\approx$ 1.0 trillion points/second, consistent with the throughput benchmark (Sec.~\ref{subsect:performance-results}).

\section{Conclusions}
\label{sect:conclusions}

We developed mach, an open-source Python beamformer with a highly optimized CUDA kernel that processes 1.1 trillion points per second on a consumer-grade GPU.
Its hybrid delay computation and memory-coalesced CUDA implementation provide $>$10$\times$ performance improvement over existing GPU beamformers while maintaining numerical accuracy.

This throughput enables real-time 3D ultrafast ultrasound reconstruction for the first time on widely available hardware.
By eliminating the beamforming bottleneck, mach enables real-time volumetric functional neuroimaging, intraoperative guidance, and other applications requiring immediate feedback from 3D ultrafast acquisitions.

mach is freely available at \url{https://github.com/Forest-Neurotech/mach} and installable via \href{https://pypi.org/project/mach-beamform/}{PyPI}.

\subsection*{Disclosures}
The authors declare that there are no financial interests, commercial affiliations, or other potential conflicts of interest that could have influenced the objectivity of this research or the writing of this paper.

\subsection*{Code, Data, and Materials Availability}
mach is openly available on GitHub at \url{https://github.com/Forest-Neurotech/mach} and can be installed from PyPI at \url{https://pypi.org/project/mach-beamform/}.
Comprehensive documentation, API references, and interactive tutorials are available at the GitHub repository.
Example code and datasets used in the benchmarks are available in the mach repository.
The PyMUST Rotating Disk dataset and PICMUS datasets used for validation are publicly available from their respective sources.

\subsection*{Acknowledgments}
This package was developed at Forest Neurotech, a Focused Research Organization supported by Convergent Research and generous philanthropic funders.
The examples and benchmarks are built on the foundation of the ultrasound open-source community, particularly vbeam, PyMUST, and PICMUS.
We thank Gustavo Zago Canal for contributing an example, Gevorg Chepchyan for prototyping CUDA optimization, Qi Zheng for guidance on CUDA profiling, and Renee Wang for illustrating Fig.~\ref{fig:transmit_receive_probe}.
Claude Sonnet was used to check language and grammar.

A previous, reduced version of this manuscript was accepted for presentation at the SPIE Medical Imaging Ultrasonic Imaging and Tomography Conference 2026 and will be included in the Medical Imaging 2026 proceedings~\cite{mach_MI2026}.


\bibliography{report}   
\bibliographystyle{spiejour}   

\subsection*{Author Biographies}

\vspace{2ex}\noindent\textbf{Charles Guan} was a Staff Data Scientist at Forest Neurotech, where he helped develop mach and Forest 1, a research ultrasound system for functional neuroimaging and neuromodulation. He completed his PhD in Bioengineering at Caltech. His interests include open-source software, brain-computer interfaces, and signal processing.

\vspace{2ex}\noindent\textbf{Alexander P. Rockhill} was a Clinical Research Data Scientist at Forest Neurotech, where he analyzed functional ultrasound imaging data. He did his PhD at the University of Oregon on intracranial electrophysiology of movement. He is currently a computational neuroscientist at FIND Neuro. Alexander is a regular contributor and maintainer of open-source software including MNE-Python, hmmlearn, and the Brain Imaging Data Structure (BIDS) standard.

\vspace{2ex}\noindent\textbf{Masashi Sode} is a Ph.D. student in the Lampe Joint Department of Biomedical Engineering at the University of North Carolina at Chapel Hill and North Carolina State University, NC, USA. His research focuses on nonlinear acoustic simulations and deep learning methods for quantifying tissue mechanical properties from ultrasound signals. He earned his M.S. in aerospace engineering from Tohoku University in Japan in 2019. Before joining UNC-Chapel Hill and NC State University, he worked as an AI engineer at a biomedical engineering company where he developed deep learning models for medical applications.

\vspace{2ex}\noindent\textbf{Gianmarco Pinton} is an Associate Professor in the Joint Department of Biomedical Engineering at the University of North Carolina at Chapel Hill and North Carolina State University. His research interests are in non-conventional diagnostic ultrasound imaging, nonlinear wave propagation, and shear shock waves and brain injury.

\listoffigures
\listoftables

\end{document}